\documentclass[twoside]{article}
\usepackage{fleqn,espcrc2}

\usepackage{graphicx}

\newcommand{\AmS}{{\protect\the\textfont2
  A\kern-.1667em\lower.5ex\hbox{M}\kern-.125emS}}

\title{Inverse Beta Decay in a Nonequilibrium Antineutrino Flux
from a Nuclear Reactor}

\author{V.I. Kopeikin\address[MCSD]{Institute for General and Nuclear
Physics, RRC "Kurchatov Institue", Moscow, Russia.},
        L.A. Mikaelyan\addressmark[MCSD],
        and
        V.V. Sinev\addressmark[MCSD].}
       
\begin{document}

\begin{abstract}
The evolution of the reactor antineutrino spectrum toward equilibrium 
above the inverse beta-decay threshold during the reactor operating period and the decay of residual 
$\bar{{\nu}_{e}}$, radiation after reactor shutdown are considered. It is found that, under certain 
conditions, these processes can play a significant role in experiments seeking neutrino oscillations. 
\vspace{1pc}
\end{abstract}

\maketitle

\section{INTRODUCTION}

The flux and the spectrum of reactor antineutrinos depend not only 
on the current reactor state, which is specified by the power level 
and the isotope composition of the nuclear fuel used, but also on
the preceding evolution of the fuel. The spectrum of antineutrinos 
produced in the beta decay of fission products and other radioactive 
nuclei accumulated in the reactor core begin to evolve toward 
equilibrium after the start-up of a reactor. After reactor shutdown, 
the antineutrino radiation diminishes for a long time.

In [1, 2], equilibration of the spectrum and the reduction of the 
soft section of the $\bar{{\nu}_{e}}$ 
spectrum, where the effects of preceding evolution are pronounced, 
were analyzed in connection with the problem of
searches for the neutrino magnetic moment in experiments studying 
$\bar{{\nu}_{e}},e^{-}$ scattering.

In this study, we consider the inverse-beta-decay process 
\begin{equation}
 \bar{{\nu}_{e}} + p \rightarrow n + e^{+} 
\end{equation}
and antineutrino spectrum above the threshold for this reaction 
(1.804 MeV). Presently, reaction (1) is of interest mainly as a tool 
for seeking neutrino neutrino oscillations in reactor experiments.

In this case, the oscillations in question are manifested in the 
disappearance of some fraction of the
$\bar{{\nu}_{e}}$ flux and in a characteristic modulation of the 
$\bar{{\nu}_{e}}$ spectrum, the latter being specified by the factor
\begin{equation}
P = 1 - {\sin}^{2}2{\theta}{\sin}^{2}(1.27{\delta}m^{2}R/E),
\end{equation}
where ${\sin}^{2}2{\theta}$ is the mixing parameter, ${\delta}m^{2}$ 
is the mass parameter measured in electronvolts squared,
$R$ is the distance (in meters) between the source and detector, 
and $E$ is the detected neutrino energy measured in megaelectronvolts. 

In experiments seeking the above oscillations, the
$\bar{{\nu}_{e}}$ spectrum and flux measured with the aid of reaction 
(1) are compared with their values 
expected in the absence of the oscillations. In this case, use is 
made of the reactor $\bar{{\nu}_{e}}$ 
spectrum obtained independently
($\bar{{\nu}_{e}}$ spectrum at the production instant). 
Uncertainties in determining this spectrum restrict the sensitivity
of the method, and systematic error in the spectrum shape may 
generally mimic or mask the oscillation
effect. Procedures that are applied to analyze data from experiments 
seeking oscillations were described
in more detail elsewhere (see, for example, [3]). As a rule, 
equilibration of the spectrum and the reduction
of the $\bar{{\nu}_{e}}$ spectrum do not lead to considerable 
effects in the region $E > 1.8 MeV$. However, their role can be
greatly enhanced in some cases that will be discussed in Section 4.

This article is organized as follows. In Section 2, we give a brief 
account of available data on the spectrum 
of reactor $\bar{{\nu}_{e}}$ in the region $E > 1.8 MeV$ and on the 
cross section for reaction (1). 
In Section 3, we consider the evolution of the spectrum in the 
reactor operating period and residual antineutrino radiation
in the reactor shutdown period and determine the relevant cross 
sections. Section 4 is devoted to discussing the results.

\section{SPECTRA AND CROSS SECTIONS
(STANDARD APPROACH}

Let us briefly review the basic properties of the
reactor-antineutrino spectrum in the energy region
above 1.8 MeV and the spectrum-averaged cross
section for reaction (1) (see, for example, [1, 2, 4-8]
and references therein).

The standard approach is based on the following
assumptions:

\begin{table*}[htb]
\caption{Yields of some fission fragments (in \%).}
\label{table:1}
\newcommand{\m}{\hphantom{$-$}}
\newcommand{\cc}[1]{\multicolumn{1}{c}{#1}}
\renewcommand{\tabcolsep}{2pc} 
\renewcommand{\arraystretch}{1.2} 
\begin{tabular}{c|c|c|c|c}
\hline
Fission  & Multicolumn{4}{Fissile nucleus} \\
\cline {2-5}
fragment & $^{235}U$ & $^{239}Pu$ & $^{241}Pu$ & $^{238}U$\\
\hline
$^{97}Zr$  & 5.95 & 5.30 & 4.89 & 5.50 \\
$^{132}I$  & 4.30 & 5.40 & 4.14 & 5.16 \\
$^{93}Y$   & 6.40 & 3.89 & 3.15 & 4.97 \\
$^{106}Ru$ & 0.40 & 4.31 & 6.18 & 2.55 \\
$^{144}Ce$ & 5.48 & 3.74 & 4.39 & 4.50 \\
$^{90}Sr$  & 5.82 & 2.10 & 1.57 & 3.12 \\
\hline
\end{tabular}\\[2pt]
\end{table*}

(i) The $\bar{{\nu}_{e}}$ spectrum is formed exclusively by the
beta decays of the fragments produced in a reactor
upon the fission of $^{235}U, ^{239}Pu, ^{238}U$ and $^{241}Pu$
isotopes.

(ii) For each isotope, the equilibrium spectrum
is established within a small time interval that can
be neglected. At the instant under consideration,
the reactor-antineutrino spectrum ${\rho}(E)$ measured in
(MeV)$^{-1}$ units per fission event can then be expressed
in terms of the spectra ${\rho}(E)$ for four fissile isotopes.
Specifically, we have
\begin{equation}
{\rho}(E) = \sum{{\alpha}_{i}{\rho}_{i}(E)}
\end{equation}
where ${\alpha}_{i}$ is the contribution of a given isotope to the
number of fission events occurring in a reactor at
this instant, the subscript values of $i = 5, 9, 8$, and
1 labeling the quantities associated with the fissile
isotopes $^{235}U$, $^{239}Pu$, $^{238}U$ and $^{241}Pu$ 

Because of $^{235}U$ depletion and the accumulation
of fissile plutonium isotopes, the contributions ${\alpha}_{i}$ 
entering into Eq. (3) change, which leads to a change
in the total spectrum ${\rho}(E)$ over the reactor operating
period. As was mentioned above, the spectra ${\rho}_{i}(E)$
are assumed to be time-independent. Information
about current ${\alpha}_{i}$ values is presented by the reactor
personnel, the relative error in these values being set
to 5\%.

The spectra ${\rho}_{i}(E)$ for $^{235}U$, $^{239}Pu$ and $^{241}Pu$
were obtained in [5] by the conversion method, which
enables one to reconstruct the relevant antineutrino
spectrum on the basis of the total spectrum of beta-
decay electrons from the set of the fragments of a
given fissile isotope. For a few tens of hours, thin
layers of the aforementioned isotopes were exposed to
a thermal-neutron flux from the reactor installed at
ILL (Grenoble), and the current spectra of electrons
from the beta decay of fragments were simultaneously
measured in the energy region above 2.0 MeV for
$^{235}U$  and in the energy region above 1.8 MeV for
$^{239}Pu$  and $^{241}Pu$ . After approximately 12 hours of 
irradiation, the beta spectra reached saturation and
changed ii longer. Therefore, the antineutrino spectra  
${\rho}_{i}(E)$ reconstructed on the basis of these $\beta$ spectra
are actually those that are established after about one
day of fuel irradiation. For $^{238}U$, use was made of the
spectrum calculated in [6] because the beta spectrum
was not measured for the fragments of E1 isotope.

The expected number of events of the interaction between 
antineutrinos and target protons was
calculated in terms of the cross section ${\sigma}_{V-A}$ for
reaction (1),
\begin{equation}
{\sigma}_{V-A} = \sum{{\alpha}_{i}{\sigma}_{i}}
\end{equation}
where ${\sigma}_{i} = \int{{\rho}_{i}(E){\sigma}(E)dE}, 
{\sigma}(E)$ being the reaction
cross section for monoenergetic antineutrinos [7].

In terms of $10^{-43}$ cm$^2$ units per fission event, the
${\sigma}_{i}$ values calculated in this way are
\begin{eqnarray}
{\sigma}_{5} = 6.39 \pm 1.9\%, {\sigma}_{9} = 4.18 \pm 2.4\% \nonumber \\
{\sigma}_{1} = 5.76 \pm 2.1\%, {\sigma}_{8} = 8.88 \pm 10\% 
\end{eqnarray}
These calculations were performed with the9 constants corresponding 
to the free-neutron lifetime of ${\tau} = 887.4 s \pm 0.2\%$. 
The resulting error in the cross
section (4) is 2.7\% (68\% C.L.).

To a higher precision, the cross section for reaction
(1) is known from an experiment performed by the
Kurchatov Institute - College de France - LAPP collaboration at a 
distance of 15 m from the Bugey PWR reactor [8]:
\begin{eqnarray}
{\sigma}_{meas} = 5.750 \times 10^{-43}cm^{2}/(fission event) \nonumber \\
         \pm 1.4\%(68\% C.L.) 
\end{eqnarray}
This result corresponds to the following contributions a; from fissile isotopes:
\begin{eqnarray}
{\alpha}_{5} = 0.538, {\alpha}_{9} = 0.328, {\alpha}_{8} = 0.078, \nonumber \\
   {\alpha}_{1} = 0.056.
\end{eqnarray}

The proportion in (7) for the number of isotope-fission events is 
typical for PWR reactors, which
were used in the majority of experiments seeking the
oscillations in question. Within the errors, the cross
section in (6) agrees with the reaction cross section
${\sigma}_{V-A}$ found for the given composition of nuclear fuel:
\begin{eqnarray}
{\sigma}_{meas}/{\sigma}_{V-A} = 0.987 \pm 1.4\%(experiment) \nonumber \\
     \pm 2.7\% (V-A).
\end{eqnarray}

The experimental value in (6) is treated as a metrological reference for 
the cross section in the absence
of oscillations. A feature peculiar to this reference is
that, in each specific case of its application, it must be
rescaled to the relevant composition of nuclear fuel.
As a result, the error in the cross section increases to about 1.6\%.

For reactors of the type being considered, the duration of the operating 
period is approximately one year. After that, the reactor is shut 
down for 30-40 days, and one-third of the fuel is replaced by a new
load of fuel. Therefore, the fuel is irradiated for three
years.

The detector background is measured during reactor shutdown periods. 
It is assumed that antineutrino emission in the region $E > 1.8$ MeV ceases
completely within one day after reaction shutdown.

The above information forms a basis for analyzing and interpreting 
the results of experiments seeking neutrino oscillations in reactor 
experiments.

\section{SPECTRA AND CROSS SECTIONS (INCLUSION OF NONEQUILIBRIUM EFFECTS)}

1. Let us determine more precisely sources that contribute to the 
formation of the reactor-antineutrino spectrum ${\rho}(E)$,
\begin{equation}
{\rho}(E) = ^{F}{\rho}(E) + ^{C}{\rho}(E)
\end{equation}

The first term in (9) describes radiation from the set of fragments 
produced in $^{235}U, ^{239}Pu, ^{238}U$ and $^{241}Pu$ fission, 
their interactions with neutrons being
disregarded here; the second term takes into account additional radiation 
arising in radiative neutron capture by accumulated fragments. 

For an irradiation time of $t_{on} > 1$ day, each of the spectra ${\rho}(E,t_{on})$ 
for the four isotopes $^{235}U, ^{239}Pu, ^{238}U$ and $^{241}Pu$
were calculated as functions of the duration of fuel irradiation. 
In this calculation, the fission rate was assumed to be constant for each
isotope. The residual-radiation spectra ${\rho}(E,t_{off})$ were
also calculated as functions of the time $t_{off}$ from the
end of fission.

The calculations were performed by summing, at each instant, the 
contributions from individual fission fragments with allowance 
for their yields from the fission process, decay diagrams, 
and lifetimes. The database that we used contains information about
571 fission fragments whose cumulative yields are not less than 
$10^{-4}\%$ each. As a matter of fact, the activity of the majority 
of the fragments that can contribute in the energy region $E > 1.8$ MeV
reaches saturation within one day after the onset of the fission process, 
and a further increase is due to only six fission products whose properties 
are well known. Three of them - $^{97}Zr$ ($E_{max} = 1.922$ MeV) $^{132}I$
($E_{max} = 2.140$ MeV), and $^{93}Y$ ($E_{max} = 2.890$ MeV) - attain
equilibrium within ten days. A further slow increase is determined by 
the $^{106}Ru$ and $^{144}Ce$ half-lives (see (10)).

\begin{table*}
\begin{eqnarray}
^{106}Ru \frac{T_{1/2}=372 day}{E_{max}=0.04 MeV} \rightarrow Rh 
\frac{T_{1/2}=30 s}{E_{max}=3.541 MeV}\rightarrow Pd(stab) \nonumber \\
^{144}Ce \frac{T_{1/2}=285 day}{E_{max}=0.32 MeV}\rightarrow Pr 
\frac{T_{1/2}=17 min day}{E_{max}=2.996 MeV}\rightarrow Nd(T_{1/2}=3\times10^{-15})yr.
\end{eqnarray}
\end{table*}

Finally, some contribution comes from $^{90}Y(T_{1/2} = 64$ h, 
$E_{max}=2.279$ MeV) as well, which is in equilibrium with its 
very long-lived predecessor $^{90}Sr$ ($T_{1/2} = 28.6$ yr). 
The yields of these fission fragments are quoted in the table. 

2. The antineutrino spectra calculated for $^{235}U$ and $^{239}Pu$ 
are displayed in Figs. la and lb (equilibration) and in Figs. 2a and 2b 
(decrease). For four fissile isotopes, Figs. 3 and 4 show, respectively, the
cross sections ${\sigma}_(t_{on})$ for reaction (1) in the fission
process and the decrease in ${\sigma}_(t_{off})$ in the residual-radiation 
spectra. In Figs. 1-4, the spectra and cross
sections are presented in dimensionless units and are
normalized to the corresponding values after a lapse
of $t_{on}$ = 1 day from the onset of the fission process.

\begin{figure}[htb]
\includegraphics[width=75mm]{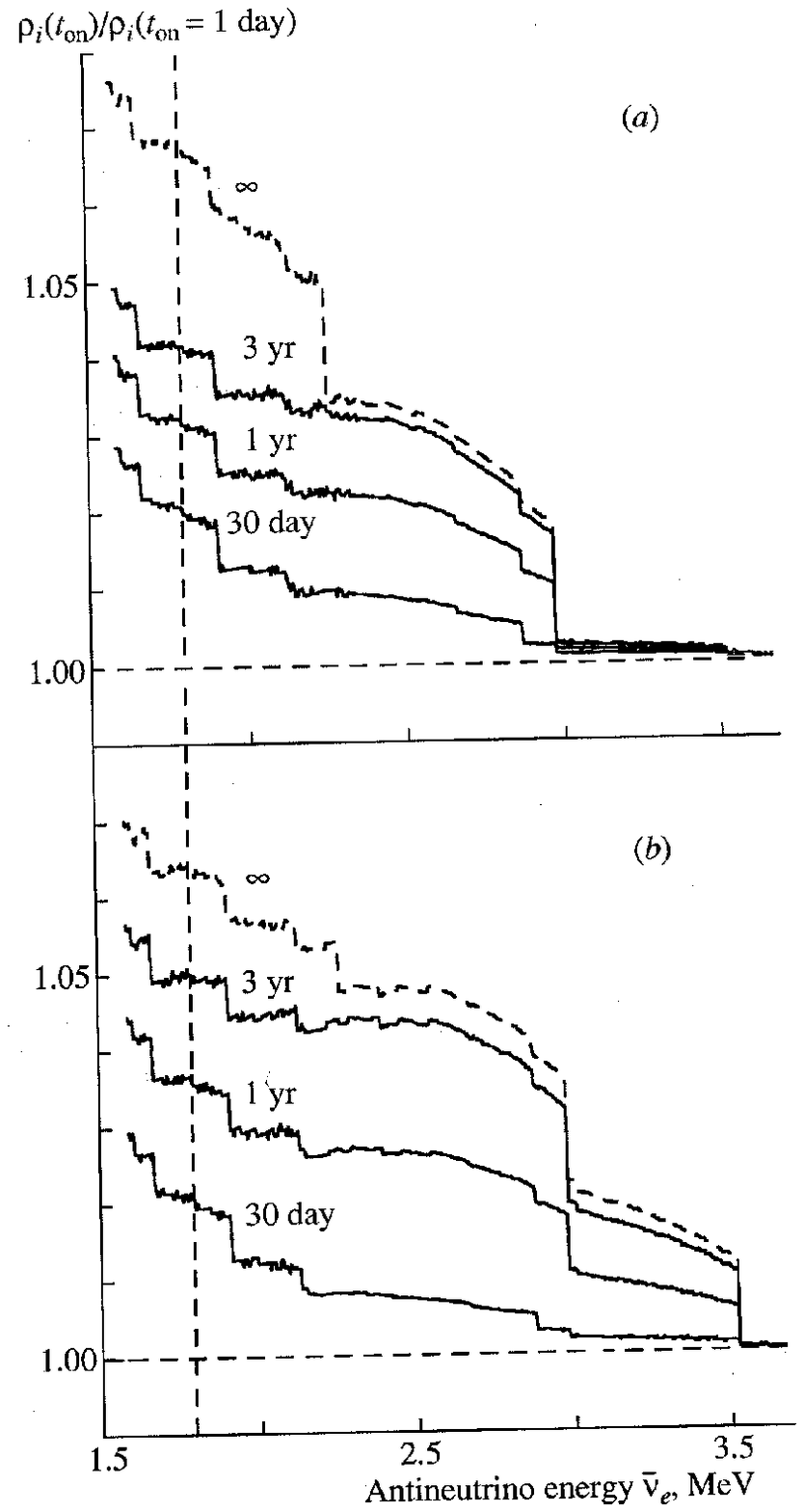}
\caption{Ratio of the antineutrino spectra for (a) $^{235}U$ and (b)
$^{239}Pu$ fission to the spectrum after one day of irradiation. The
numbers on the curves indicate the irradiation time.}
\label{fig:f1}
\end{figure}

\begin{figure}[htb]
\includegraphics[width=75mm]{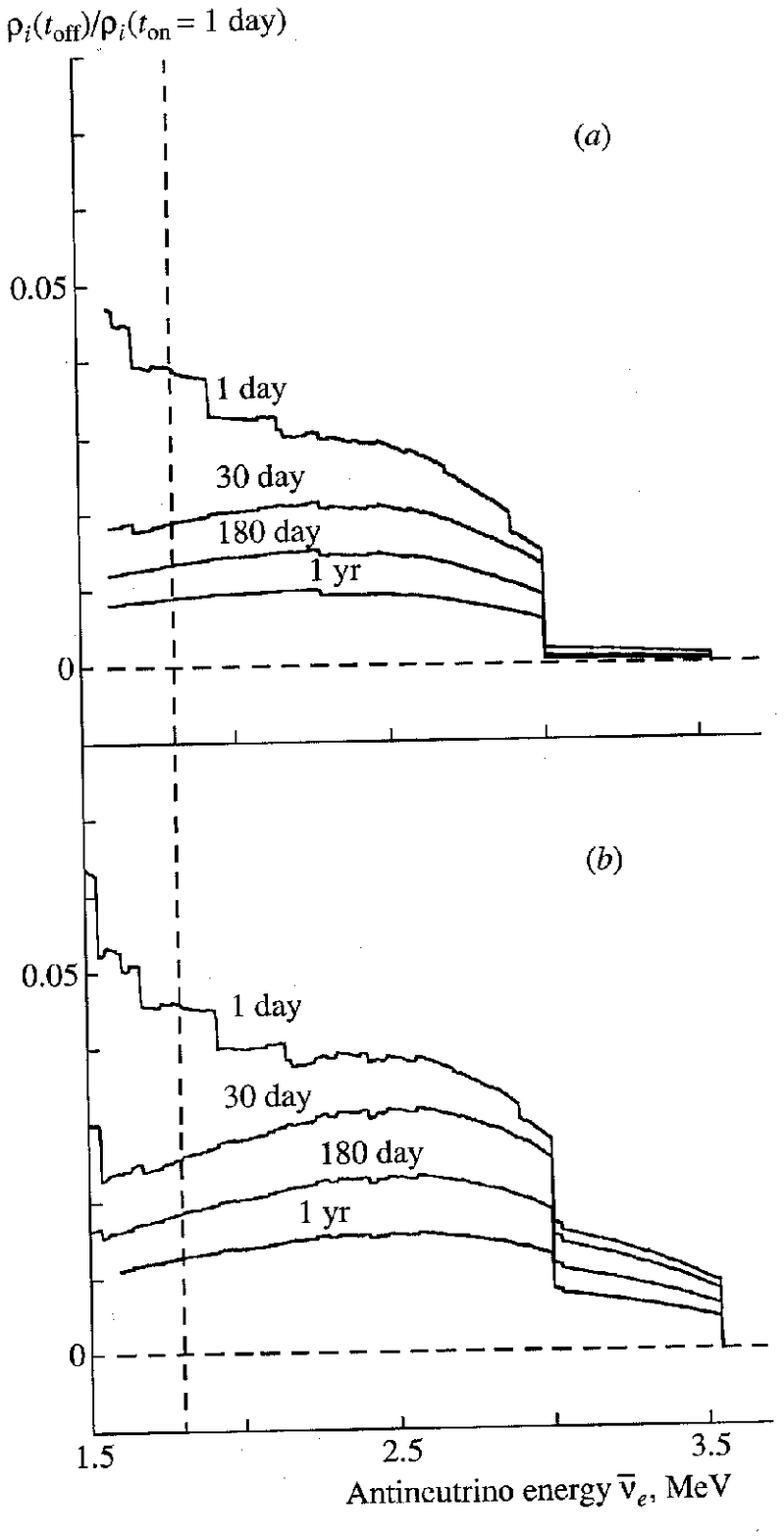}
\caption{Ratio of the antineutrino spectra of the residiul radiation 
from (a) $^{235}U$ and (b)$^{239}Pu$ irradiated for two years
to the spectrum after one day of irradiation. The numbers on the 
curves indicate the time that  have elapsed after the termiation
of irradiation.}
\label{fig:f2}
\end{figure}

3. The correction ${\rho}(E)$ was calculated with allowance for fragments 
accumulated in a reactor, the cross sections for radiative neutron capture, 
and the spatial and energy distributions of the neutron flux over the 
reactor core. The results presented in Fig. 5 refer to a standard operating 
period of a PWR reactor. 

The relevant contribution $^{C}{\rho}(E)$ to the cross section for reaction 
(1) is about 0.2\% at the end of the operating period. After reactor 
shutdown, the spectrum $^{C}{\rho}(E)$ decreases fast, not making a significant 
contribution to residual radiation. 4. 

\section{DISCUSSION OF RESULTS}

(i) In the approximation specified in Section 2, the flux and the spectrum 
of reactor antineutrinos are un- ambiguously determined by the current reactor 
state. After sharp changes in this state, the characteristics of the $\bar{{\nu}_{e}}$
flux take the corresponding equilibrium values within a time as short as 
one day. In particular, the antineutrino flux falls down to zero within one day
after reactor shutdown.

\begin{figure}[htb]
\includegraphics[width=75mm]{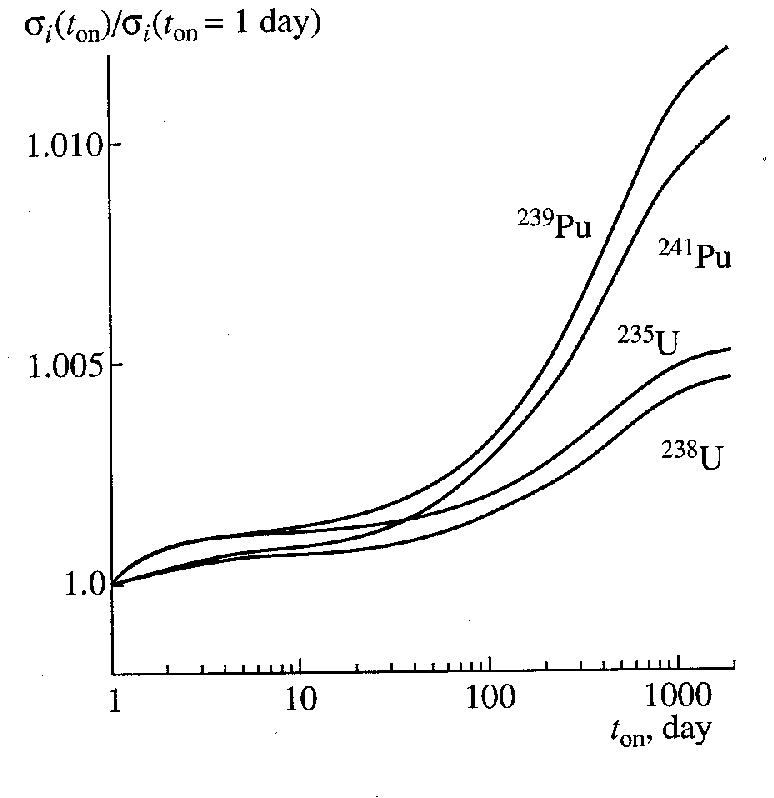}
\caption{Ratio of the inverse-beta-decay cross sections for
fissile uranium and plutonium isotopes irradiated for the
time $t_{on}$ to that after one day of irradiation.}
\label{fig:f3}
\end{figure}

\begin{figure}[htb]
\includegraphics[width=75mm]{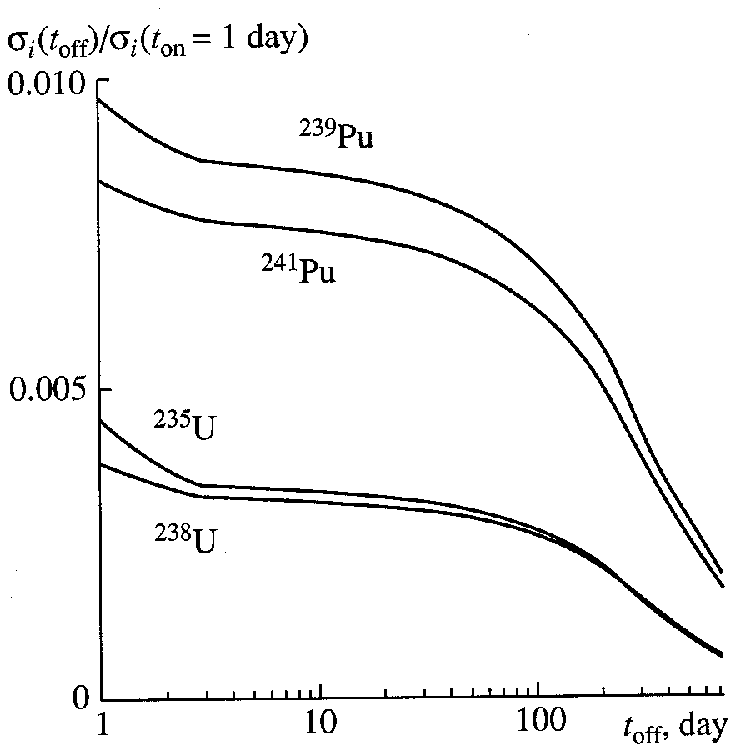}
\caption{Ratio of the cross sections for inverse beta decay
induced by the residual-radiation antineutrinos for uranium and 
plutonium isotopes irradiated for two years to
the cross section after one day of irradiation as a function
of the time $t_{off}$ after the termination of irradiation.}
\label{fig:f4}
\end{figure}

\begin{figure}[htb]
\includegraphics[width=75mm]{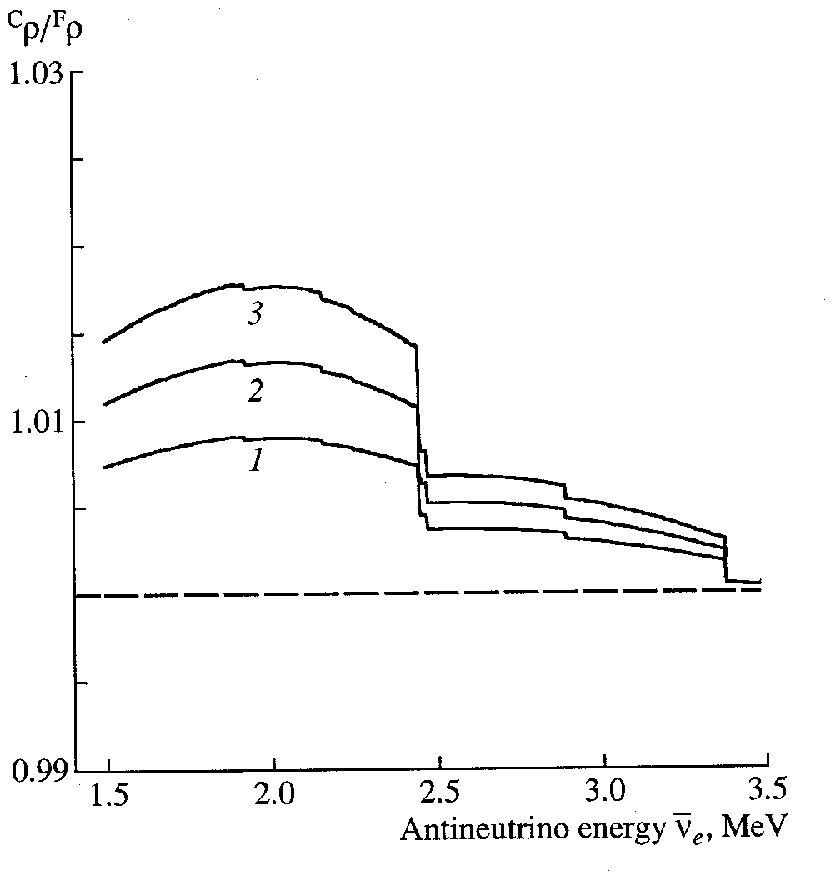}
\caption{Ratio of the cross sections for inverse beta decay
induced by the residual-radiation antineutrinos for uranium and 
plutonium isotopes irradiated for two years to
the cross section after one day of irradiation as a function
of the time $t_{off}$ after the termination of irradiation.}
\label{fig:f5}
\end{figure}

As a matter of fact, the antineutrino flux in the region $E > 1.8$
MeV has been found to have a nonequilibrium component, whose
relaxation time exceeds the duration of the reactor operation period.
In a view of this, a determaination of the $\bar{{\nu}_{e}}$ spectrum
and of the corresponding values of the cross section for reaction (1)
requires tracing the preceding evolution of reactor operation over a long
time and taking into account power levels, shutdown periods and the 
discharge of spent reactor fuel. In each specific case, this can be done, 
if needed to a sufficient precision.

The sample of the results in Section 3, which were obtained under 
assumption that the fission of uranium and plutonium isotopes
proceeds at a constant rate, has enabled us to reveal qualitative features
of effects induced by the nonequilibrium component.

(ii) First of all, we note that the resulting corrections to the spectrum
and cross sections are relatively small, but they are not negligible. In the 
antineutrino energy range 1.8-3.5 MeV, the relative contribution of the
additional radiation during the reactor operating period (Figs. 1, 5) is 
about 4\%, which is somewhat greater than the error of the ILL spectra [5].

The corrections to the cross sections cr, in Fig. 3 and the correction 
associated with radiative neutron capture may change the cross sections 
${\sigma}_{V-A}$ by 0.4-0.6\%. As a result, relation (8) will change 
accordingly. Corrections on the same order of magnitude
may arise if the cross section (6) is used as a reference value for 
the cross section in the absence of the oscillations.

(iii) Here, we consider a situation where residual radiation from a stopped 
reactor can play a significant role and provide numerical examples 
illustrating the scale of the effects under discussion.

Let us consider an experiment where antineutrinos from reactors 
are recorded by one detector positioned in such a way that the reactors 
are at markedly different distances from the detector. Such an 
experimental setting was implemented, for example, in Rovno
(distances of 18 and 98 m) and in the Bugey-3 experiment (15 and 95 m) 
[9]; three reactors in Krasnoyarsk were located at distances of 57, 57, 
and 234 m from the detector [10]. The idea of these experiments consists 
in using the shutdown and operating periods to determine the background 
level and the signals from each individual reactor. By comparing these 
signals, one can reveal the oscillation effect or set limits on the
oscillation parameters.

By way of example, we consider two identical PWR reactors of thermal power 
2.8 GW each and a detector positioned at distances of 15 and 100 m
from the near and the far reactor, respectively, so 

Fig.5. Ratio of the spectrum component ?(A) 
associated with neutron capture by fission products in a PWR reactor to the 
fission spectrum p(E). Curves /, 2, and 8 correspond to the beginning, middle, 
and end of the reactor operating period. 

that the signal from the near reactor is approximately 45 times as great 
as that from the far one. In this case, the numbers of events of reaction 
(1) that are induced by the near and the far reactor per 1 t of a
$CH_{1.8}$ target per day are $N^{op}_{near} \approx 12000$ day$^{-1}$ t$^{-1}$
and $N^{op}_{far} \approx 260$ day$^{-1}$ t$^{-1}$ ,respectively.

When both reactors are shut down, the detector records, however, the 
residual interaction in addition to the background. Within one to 
two days after the shutdown of the near reactor, the number of
interaction events induced by the residual radiation
from it in the detector per unit time decreases by a
factor of about 200, falling down to a value of 
$N^{off}_{near} \approx 70$ day$^{-1}$ t$^{-1}$, which is approximately 25\% 
of the signal from the far reactor. After that, the signal from
the residual radiation decreases smoothly (see Fig. 4)
until the discharge of the spent fuel begins, which
significantly affects detector readings.

Obviously, these cases require a new approach to processing and analyzing 
experimental data.

\section{CONCLUSION}

The effect of the equilibration of the reactor-antineutrino spectrum on 
the cross section for the inverse-beta-decay process (1) with the 
threshold of 1.8 MeV has been considered.
The increase in the antineutrino flux due to an increase in the cross 
section (this effect was previously ignored) is about 0.6\% over 
the reactor operating period. This value is commensurate with the accuracy
(the standard error is 1.4\%) in the measurement of the cross section 
itself. A correction at this level must also be introduced in the ratio 
of the measured cross section to the cross section calculated on the basis
of the antineutrino spectrum that was determined independently and 
which corresponds to one day of reactor operation.

We have calculated the additional contribution to
the antineutrino spectrum from neutron capture by
fission products and determined the corresponding
increase (0.2\%) in the cross section for reaction (1). It has been found that 
corrections previously ignored in the standard experimental scheme aimed at 
searches for neutrino oscillations (one detector and near and far reactors) may 
be as large as 25\%. 

\section*{ACKNOWLEDGMENTS}
We are grateful to Prof. K. Schreckenbach for placing valuable 
information at our disposal and to our colleagues participating in the CHOOZ 
collaboration for stimulating discussions. This work was supported by the 
Russian Foundation for Basic Research (project nos. 00-15-06708 and 00-02-16035).

\end{document}